\documentstyle[aps,preprint]{revtex}

\begin{document}

\title{SPACETIME FOAM AND THE CASIMIR ENERGY}
\author{Remo Garattini$^{1,2}$ \\
(1) M\'{e}canique et Gravitation, Universit\'{e} de\\
Mons-Hainaut, Facult\'{e} des Sciences, 15 Avenue\\
Maistriau, B-7000 Mons, Belgium.\\
{(2) Universit\'{a} degli Studi di Bergamo} Dalmine\\
(Bergamo), 24044, Italy. E-mail: Garattini@mi.infn.it}
\maketitle

\begin{abstract}
We conjecture that the neutral black hole pair production is related to the
vacuum fluctuation of pure gravity via the Casimir-like energy. A
generalization of this process to a multi-black hole pair is considered.
Implications on the foam-like structure of spacetime and on the cosmological
constant are discussed.
\end{abstract}

\section{Introduction}

It was J. A. Wheeler who first conjectured that spacetime could be subjected
to topology fluctuation at the Planck scale\cite{Wheeler}. This means that
spacetime undergoes a deep and rapid transformation in its structure. This
changing spacetime is best known as ``{\it spacetime foam}'', which can be
taken as a model for the quantum gravitational vacuum. Some authors have
investigated the effects of such a {\it foamy} space on the cosmological
constant, one example is the celebrated Coleman mechanism involving
wormholes \cite{Coleman}. Nevertheless, how to realize such a foam-like
space is still unknown as too is whether this represents the real quantum
gravitational vacuum. For this purpose, we begin to consider the
``simplest'' quantum process that could approximate the foam structure in
absence of matter fields, that is the black hole pair creation. Different
examples are known on this subject. The first example involves the study of
black hole pair creation in a background magnetic field represented by the
Ernst solution \cite{Ernst} which asymptotically approaches the Melvin
metric \cite{Melvin}. Another example is the Schwarzschild-deSitter metric
(SdS) which asymptotically approaches the deSitter metric. The extreme
version is best known as the Nariai metric\cite{Nariai}. In this case the
background is represented by the cosmological constant $\Lambda $ acting on
the neutral black hole pair produced, accelerating the components away from
each other. Finally another example is given by the Schwarzschild metric
which asymptotically approaches the flat metric and depends only on the mass
parameter $M$. Metrics of this type are termed asymptotically flat (A.F.).
Another metric which has the property of being A.F. is the
Reissner-Nordstr\"{o}m metric, which depends on two parameters: the mass $M$
and the charge $Q$ of the electromagnetic field. Nevertheless, all the cases
mentioned above introduce an external background field like the magnetic
field or the cosmological constant to produce the pair and accelerate the
components far away. In this letter, we would like to consider the same
process without the contribution of external fields, except gravity itself
and consider the possible implications on the foam-like structure of
spacetime. This choice linked to the vacuum Einstein's equations leads to
the Schwarzschild and the flat metrics, where only the simplest case is
considered, i.e. metrics which are spherically symmetric. Since the A.F.
spacetimes are non-compact a subtraction scheme is needed to recover the
correct equations under the constraint of fixed induced metrics on the
boundary\cite{Frolov,HawHor}.

\section{Quasilocal Energy and Entropy in presence of a Bifurcation Surface}

Although it is not necessary for the forthcoming discussions, let us
consider the maximal analytic extension of the Schwarzschild metric, i.e.,
the Kruskal manifold whose spatial slices $\Sigma $ represent Einstein-Rosen
bridges with wormhole topology $S^{2}\times R^{1}$. Following Ref.\cite
{Frolov}, the complete manifold ${\cal M}$ can be taken as a model for an
eternal black hole composed of two wedges ${\cal M}_{+}$ and ${\cal M}_{-}$
located in the right and left sectors of a Kruskal diagram. The hypersurface
$\Sigma $ is divided in two parts $\Sigma _{+}$ and $\Sigma _{-}$ by a
bifurcation two-surface $S_{0}$. On $\Sigma $ we can write the gravitational
Hamiltonian
\[
H_{p}=H-H_{0}=\frac{1}{2\kappa }\int_{\Sigma }d^{3}x(N{\cal H+}N^{i}{\cal H}%
_{i})
\]
\begin{equation}
+\text{ }\frac{1}{\kappa }\int_{S_{+}}^{{}}d^{2}xN\sqrt{\sigma }\left(
k-k^{0}\right) -\frac{1}{\kappa }\int_{S_{-}}d^{2}xN\sqrt{\sigma }\left(
k-k^{0}\right) ,  \label{a1}
\end{equation}
where $\kappa =8\pi G$. The Hamiltonian has both volume and boundary
contributions. The volume part involves the Hamiltonian and momentum
constraints
\[
{\cal H}=\left( 2\kappa \right) G_{ijkl}\pi ^{ij}\pi ^{kl}-\sqrt{^{3}g}%
R/\left( 2\kappa \right) =0,
\]
\begin{equation}
{\cal H}_{i}=-2\pi _{i|j}^{j}=0,
\end{equation}
where $G_{ijkl}=\left( g_{ik}g_{jl}+g_{il}g_{jk}-g_{ij}g_{kl}\right) /\left(
2\sqrt{g}\right) $ and $R$ denotes the scalar curvature of the surface $%
\Sigma $. The volume part of the Hamiltonian (\ref{a1}) is zero when the
Hamiltonian and momentum constraints are imposed. However, for the flat and
the Schwarzschild space, constraints are immediately satisfied, then in this
context the total Hamiltonian reduces to
\begin{equation}
H_{p}=\frac{1}{\kappa }\int_{S_{+}}^{{}}d^{2}xN\sqrt{\sigma }\left(
k-k^{0}\right) -\frac{1}{\kappa }\int_{S_{-}}d^{2}xN\sqrt{\sigma }\left(
k-k^{0}\right) .  \label{a1a}
\end{equation}
Quasilocal energy is defined as the value of the Hamiltonian that generates
unit time translations orthogonal to the two-dimensional boundaries, i.e.
\[
E_{tot}=E_{+}-E_{-},
\]
\[
E_{+}=\frac{1}{\kappa }\int_{S_{+}}^{{}}d^{2}x\sqrt{\sigma }\left(
k-k^{0}\right)
\]
\begin{equation}
E_{-}=-\frac{1}{\kappa }\int_{S_{-}}d^{2}x\sqrt{\sigma }\left(
k-k^{0}\right) .
\end{equation}
where $\left| N\right| =1$ at both $S_{+}$ and $S_{-}$. $E_{tot}$ is the
quasilocal energy of a spacelike hypersurface $\Sigma =\Sigma _{+}\cup
\Sigma _{-}$ bounded by two boundaries $^{3}S_{+}$ and $^{3}S_{-}$ located
in the two disconnected regions $M_{+}$ and $M_{-}$ respectively. We have
included the subtraction terms $k^{0}$ for the energy. $k^{0}$ represents
the trace of the extrinsic curvature corresponding to embedding in the
two-dimensional boundaries $^{2}S_{+}$ and $^{2}S_{-}$ in three-dimensional
Euclidean space. Let us consider the case of the static Einstein-Rosen
bridge whose metric is defined as:
\begin{equation}
ds^{2}=-N^{2}dt^{2}+g_{yy}dy^{2}+r^{2}\left( y\right) d\Omega ^{2},
\label{a1b}
\end{equation}
where $N$, $g_{yy}$, and $r$ are functions of the radial coordinate $y$
continuously defined on ${\cal M}$, with $dy=dr/\sqrt{1-\frac{2m}{r}}$. If
we make the identification $N^{2}=1-\frac{2m}{r}$, the line element (\ref
{a1b}) reduces to the S metric written in another form. The boundaries $%
^{2}S_{+}$ and $^{2}S_{-}$ are located at coordinate values $y=y_{+}$ and $%
y=y_{-}$ respectively. The normal to the boundaries is $n^{\mu }=\left(
h^{yy}\right) ^{\frac{1}{2}}\delta _{y}^{\mu }$. Since this normal is
defined continuously along $\Sigma $, the value of $k$ depends on the
function $r,_{y}$, which is positive for $^{2}B_{+}$ and negative for $%
^{2}B_{-}$. The application of the quasilocal energy definition gives
\[
E=E_{+}-E_{-}
\]
\begin{equation}
=\left( r\left| r,_{y}\right| \left[ 1-\left( h^{yy}\right) ^{\frac{1}{2}}%
\right] \right) _{y=y_{+}}-\left( r\left| r,_{y}\right| \left[ 1-\left(
h^{yy}\right) ^{\frac{1}{2}}\right] \right) _{y=y-}.
\end{equation}
It is easy to see that $E_{+}$ and $E_{-}$ tend individually to the ${\cal %
ADM}$ mass ${\cal M}$ when the boundaries $^{3}B_{+}$ and $^{3}B_{-}$ tend
respectively to right and left spatial infinity. It should be noted that the
total energy is zero for boundary conditions symmetric with respect to the
bifurcation surface, i.e.,
\begin{equation}
E=E_{+}-E_{-}=M+\left( -M\right) =0,  \label{a2}
\end{equation}
where the asymptotic contribution has been considered. The same behaviour
appears in the entropy calculation for the physical system under
examination. Indeed
\begin{equation}
S_{tot}=S_{+}-S_{-}=\exp \left( \frac{A^{+}}{4}-\frac{A^{-}}{4}\right)
\simeq \exp \left( \frac{A_{H}}{4}-\frac{A_{H}}{4}\right) =\exp \left(
0\right) ,  \label{a3}
\end{equation}
where $A^{+}$ and $A^{-}$ have the same meaning as $E_{+}$ and $E_{-}$. Note
that for both entropy and energy this result is obtained at zero loop. We
can also see Eqs. (\ref{a2}) and (\ref{a3}) from a different point of view.
In fact these Eqs. say that flat space can be thought of as a composition of
two pieces: the former, with positive energy, in the region $\Sigma _{+}$
and the latter, with negative energy, in the region $\Sigma _{-}$, where the
positive and negative concern the bifurcation surface (hole) which is formed
due to a topology change of the manifold. The most appropriate mechanism to
explain this splitting seems to be a black hole pair creation.

\section{Black Hole Pair Creation}

The formation of neutral black hole pairs with the two holes residing in the
same universe is believed to be a highly suppressed process, at least for $%
\Lambda \gg 1$ in Planck's units\cite{Bousso-Hawking}. The metric which
describes such pair creation is the Nariai metric. When the cosmological
constant is absent the SdS metric is reduced to the Schwarzschild metric
which concerns a single black hole. However, one could regard each single
Schwarzschild black hole in our universe as a mere part of a neutral pair,
with the partner residing in the other universe. In this case the whole
spacetime can be regarded as a black-hole pair formed up by a black hole
with positive mass $M$ in the coordinate system of the observer and an {\it %
anti black-hole} with negative mass $-M$ in the system where the observer is
not present. From the instantonic point of view, one can represent neutral
black hole pairs as instantons with zero total energy. An asymptotic
observer in one universe would interpret each such pair as being formed by
one black hole with positive mass $M$. What such an observer would actually
observe from the pair is only either a black hole with positive energy or a
wormhole mouth opening to the observer's universe, interpreting that the
black hole in the ``{\it other universe}'' has negative mass without
violating the positive-energy theorems\cite{Remo,PGDiaz}. This scenario
gives spacetime a different structure. Indeed it is well known that flat
spacetime cannot spontaneously generate a black hole, otherwise energy
conservation would be violated. In other terms we cannot compare spacetimes
with different asymptotic behaviour\cite{Witten}. The different boundary
conditions reflect on the fact that flat space is not periodic in euclidean
time which means that the temperature is zero. On the other hand a black
hole with an imaginary time necessitates periodicity, but this implies a
temperature different from zero. Then, unless flat spacetime has a
temperature $T$ equal to the black hole temperature, there is no chance for
a transition from flat to curved spacetime. This transition is a decay from
the false vacuum to the true one\cite{Coleman1,Perry,Gross,Mazur}. However,
taking account a pair of neutral black holes living in different universes,
there is no decay and more important no temperature is necessary to change
from flat to curved space. This could be related with a vacuum fluctuation
of the metric which can be measured by the Casimir energy.

\section{Casimir Energy}

One can in general formally define the Casimir energy as follows
\begin{equation}
E_{Casimir}\left[ \partial {\cal M}\right] =E_{0}\left[ \partial {\cal M}%
\right] -E_{0}\left[ 0\right] ,
\end{equation}
where $E_{0}$ is the zero-point energy and $\partial {\cal M}$ is a
boundary. For zero temperature, the idea underlying the Casimir effect is to
compare vacuum energies in two physical distinct configurations. We can
recognize that the expression which defines quasilocal energy is formally of
the Casimir type. Indeed, the subtraction procedure present in Eq.(\ref{a1a}%
) describes an energy difference between two distinct situations with the
same boundary conditions. However, while the expression contained in Eq.(\ref
{a1a}) is only classical, the Casimir energy term has a quantum nature. One
way to escape from this disagreeable situation is the induced gravity point
of view discussed in Ref.\cite{B.L.} and Refs. therein. However, in those
papers the subtraction procedure in the energy term is generated by the zero
point quantum fluctuations of matter fields. Nevertheless, we are working in
the context of pure gravity, therefore quasilocal energy has to be
interpreted as the zero loop or tree level approximation to the true Casimir
energy. To this end it is useful to consider a generalized subtraction
procedure extended to the volume term up to the quadratic order. This
corresponds to the semiclassical approximation of quasilocal energy. What
are the possible effects on the foam-like scenario? Suppose we enlarge this
process from one pair to a large but fixed number of such pairs, say $N$.
What we obtain is a multiply connected spacetime with $N$ holes inside the
manifold, each of them acting as a single bifurcation surface with the sole
condition of having symmetry with respect to the bifurcation surface even at
finite distance. Let us see in which way such a spacetime can be modeled.

\section{Spacetime foam: the model}

In the one-wormhole approximation we have used an eternal black hole, to
describe a complete manifold ${\cal M}$, composed of two wedges ${\cal M}%
_{+} $ and ${\cal M}_{-}$ located in the right and left sectors of a Kruskal
diagram. The spatial slices $\Sigma $ represent Einstein-Rosen bridges with
wormhole topology $S^{2}\times R^{1}$. Also the hypersurface $\Sigma $ is
divided in two parts $\Sigma _{+}$ and $\Sigma _{-}$ by a bifurcation
two-surface $S_{0}$. We begin with the line element
\begin{equation}
ds^{2}=-N^{2}\left( r\right) dt^{2}+\frac{dr^{2}}{1-\frac{2m}{r}}%
+r^{2}\left( d\theta ^{2}+\sin ^{2}\theta d\phi ^{2}\right)
\end{equation}
and we consider the physical Hamiltonian defined on $\Sigma $%
\[
H_{P}=H-H_{0}=\frac{1}{16\pi l_{p}^{2}}\int_{\Sigma }d^{3}x\left( N{\cal H}%
+N_{i}{\cal H}^{i}\right)
\]
\begin{equation}
+\frac{2}{l_{p}^{2}}\int_{S_{+}}^{{}}d^{2}x\sqrt{\sigma }\left(
k-k^{0}\right) -\frac{2}{l_{p}^{2}}\int_{S_{-}}d^{2}x\sqrt{\sigma }\left(
k-k^{0}\right) ,
\end{equation}
where $l_{p}^{2}=G$. The volume term contains two constraints
\begin{equation}
\left\{
\begin{array}{l}
{\cal H}=G_{ijkl}\pi ^{ij}\pi ^{kl}\left( \frac{l_{p}^{2}}{\sqrt{g}}\right)
-\left( \frac{\sqrt{g}}{l_{p}^{2}}\right) R^{\left( 3\right) }=0 \\
{\cal H}^{i}=-2\pi _{|j}^{ij}=0
\end{array}
\right. ,
\end{equation}
both satisfied by the Schwarzschild and Flat metric respectively. The
supermetric is $G_{ijkl}=\frac{1}{2}\left(
g_{ik}g_{jl}+g_{il}g_{jk}-g_{ij}g_{kl}\right) $ and $R^{\left( 3\right) }$
denotes the scalar curvature of the surface $\Sigma $. By using the
expression of the trace
\begin{equation}
k=-\frac{1}{\sqrt{h}}\left( \sqrt{h}n^{\mu }\right) _{,\mu },
\end{equation}
with the normal to the boundaries defined continuously along $\Sigma $ as $%
n^{\mu }=\left( h^{yy}\right) ^{\frac{1}{2}}\delta _{y}^{\mu }$. The value
of $k$ depends on the function $r,_{y}$, where we have assumed that the
function $r,_{y}$ is positive for $S_{+}$ and negative for $S_{-}$. We
obtain at either boundary that
\begin{equation}
k=\frac{-2r,_{y}}{r}.
\end{equation}
The trace associated with the subtraction term is taken to be $k^{0}=-2/r$
for $B_{+}$ and $k^{0}=2/r$ for $B_{-}$. Then the quasilocal energy with
subtraction terms included is
\begin{equation}
E_{{\rm quasilocal}}=l_{p}^{2}\left( E_{+}-E_{-}\right) =l_{p}^{2}\left[
\left( r\left[ 1-\left| r,_{y}\right| \right] \right) _{y=y_{+}}-\left( r%
\left[ 1-\left| r,_{y}\right| \right] \right) _{y=y_{-}}\right] .
\end{equation}
Note that the total quasilocal energy is zero for boundary conditions
symmetric with respect to the bifurcation surface $S_{0}$ and this is the
necessary condition to obtain instability with respect to the flat space. In
this sector satisfy the constraint equations (\ref{a1a}). Here we consider
perturbations at $\Sigma $ of the type
\begin{equation}
g_{ij}=\bar{g}_{ij}+h_{ij},
\end{equation}
where $\bar{g}_{ij}$ is the spatial part of the Schwarzschild and Flat
background in a WKB approximation. In this framework we have computed the
quantity
\begin{equation}
\Delta E\left( M\right) =\frac{\left\langle \Psi \left|
H^{Schw.}-H^{Flat}\right| \Psi \right\rangle }{\left\langle \Psi |\Psi
\right\rangle }+\frac{\left\langle \Psi \left| H_{quasilocal}\right| \Psi
\right\rangle }{\left\langle \Psi |\Psi \right\rangle },
\end{equation}
by means of a variational approach, where the WKB functionals are
substituted with trial wave functionals. This quantity is the natural
extension to the volume term of the subtraction procedure for boundary terms
and it is interpreted as the Casimir energy related to vacuum fluctuations.
By restricting our attention to the graviton sector of the Hamiltonian
approximated to second order, hereafter referred as $H_{|2}$, we define
\[
E_{|2}=\frac{\left\langle \Psi ^{\perp }\left| H_{|2}\right| \Psi ^{\perp
}\right\rangle }{\left\langle \Psi ^{\perp }|\Psi ^{\perp }\right\rangle },
\]
where
\[
\Psi ^{\perp }=\Psi \left[ h_{ij}^{\perp }\right] ={\cal N}\exp \left\{ -%
\frac{1}{4l_{p}^{2}}\left[ \left\langle \left( g-\bar{g}\right) K^{-1}\left(
g-\bar{g}\right) \right\rangle _{x,y}^{\perp }\right] \right\} .
\]
After having functionally integrated $H_{|2}$, we get
\begin{equation}
H_{|2}=\frac{1}{4l_{p}^{2}}\int_{\Sigma }d^{3}x\sqrt{g}G^{ijkl}\left[
K^{-1\bot }\left( x,x\right) _{ijkl}+\left( \triangle _{2}\right)
_{j}^{a}K^{\bot }\left( x,x\right) _{iakl}\right]
\end{equation}
The propagator $K^{\bot }\left( x,x\right) _{iakl}$ comes from a functional
integration and it can be represented as
\begin{equation}
K^{\bot }\left( \overrightarrow{x},\overrightarrow{y}\right)
_{iakl}:=\sum_{N}\frac{h_{ia}^{\bot }\left( \overrightarrow{x}\right)
h_{kl}^{\bot }\left( \overrightarrow{y}\right) }{2\lambda _{N}\left(
p\right) },
\end{equation}
where $h_{ia}^{\bot }\left( \overrightarrow{x}\right) $ are the
eigenfunctions of
\begin{equation}
\left( \triangle _{2}\right) _{j}^{a}:=-\triangle \delta
_{j}^{a_{{}}^{{}}}+2R_{j}^{a}.
\end{equation}
This is the Lichnerowicz operator projected on $\Sigma $ acting on traceless
transverse quantum fluctuations and $\lambda _{N}\left( p\right) $ are
infinite variational parameters. $\triangle $ is the curved Laplacian
(Laplace-Beltrami operator) on a Schwarzschild background and $R_{j\text{ }%
}^{a}$ is the mixed Ricci tensor whose components are:
\begin{equation}
R_{j}^{a}=diag\left\{ \frac{-2MG}{r_{{}}^{3}},\frac{MG}{r_{{}}^{3}},\frac{MG%
}{r_{{}}^{3}}\right\} .
\end{equation}
The minimization with respect to $\lambda $ and the introduction of a high
energy cutoff $\Lambda $ give to the Eq. (\ref{a2}) the following form
\begin{equation}
\Delta E\left( M\right) \sim -\frac{V}{32\pi ^{2}}\left( \frac{3MG}{r_{0}^{3}%
}\right) ^{2}\ln \left( \frac{r_{0}^{3}\Lambda ^{2}}{3MG}\right) ,
\end{equation}
where $V$ is the volume of the system and $r_{0}$ is related to the minimum
radius compatible with the wormhole throat. We know that the classical
minimum is achieved when $r_{0}=2MG$. However, it is likely that quantum
processes come into play at short distances, where the wormhole throat is
defined, introducing a {\it quantum} radius $r_{0}>2MG$. We now compute the
minimum of $\Delta E\left( M\right) $, after having rescaled the variable $M$
to a scale variable $x=3MG/\left( r_{0}^{3}\Lambda ^{2}\right) $. Thus
\[
\Delta E\left( M\right) \rightarrow \Delta E\left( x,\Lambda \right) =\frac{V%
}{32\pi ^{2}}\Lambda ^{4}x^{2}\ln x
\]
We obtain two values for $x$: $x_{1}=0$, i.e. flat space and $x_{2}=e^{-%
\frac{1}{2}}$. At the minimum we obtain
\begin{equation}
\Delta E\left( x_{2}\right) =-\frac{V}{64\pi ^{2}}\frac{\Lambda ^{4}}{e}.
\end{equation}
Nevertheless, there exists another part of the spectrum which has to be
considered: the discrete spectrum containing one mode. This gives the energy
an imaginary contribution, namely we have discovered an unstable mode\cite
{GPY,Remo1}. Let us briefly recall, how this appears. The eigenvalue
equation
\begin{equation}
\left( \triangle _{2}\right) _{i}^{a}h_{aj}=\alpha h_{ij}
\end{equation}
can be studied with the Regge-Wheeler method. The perturbations can be
divided in odd and even components. The appearance of the unstable mode is
governed by the gravitational field component $h_{11}^{even}$. Explicitly
\[
-E^{2}H\left( r\right)
\]
\begin{equation}
=-\left( 1-\frac{2MG}{r}\right) \frac{d^{2}H\left( r\right) }{dr^{2}}+\left(
\frac{2r-3MG}{r^{2}}\right) \frac{dH\left( r\right) }{dr}-\frac{4MG}{r^{3}}%
H\left( r\right) ,  \label{a4}
\end{equation}
where
\begin{equation}
h_{11}^{even}\left( r,\vartheta ,\phi \right) =\left[ H\left( r\right)
\left( 1-\frac{2m}{r}\right) ^{-1}\right] Y_{00}\left( \vartheta ,\phi
\right)
\end{equation}
and $E^{2}>0$. Eq. (\ref{a4}) can be transformed into
\begin{equation}
\mu =\frac{\int\limits_{0}^{\bar{y}}dy\left[ \left( \frac{dh\left( y\right)
}{dy}\right) ^{2}-\frac{3}{2\rho \left( y\right) ^{3}}h\left( y\right)
\right] }{\int\limits_{0}^{\bar{y}}dyh^{2}\left( y\right) },
\end{equation}
where $\mu $ is the eigenvalue, $y$ is the proper distance from the throat
in dimensionless form. If we choose $h\left( \lambda ,y\right) =\exp \left(
-\lambda y\right) $ as a trial function we numerically obtain $\mu =-.701626$%
. In terms of the energy square we have
\begin{equation}
E^{2}=-.\,\allowbreak 175\,41/\left( MG\right) ^{2}
\end{equation}
to be compared with the value $E^{2}=-.\,\allowbreak 19/\left( MG\right)
^{2} $ of Ref.\cite{GPY}. Nevertheless, when we compute the eigenvalue as a
function of the distance $y$, we discover that in the limit $\bar{y}%
\rightarrow 0$,
\begin{equation}
\mu \equiv \ \mu \left( \lambda \right) =\lambda ^{2}-\frac{3}{2}+\frac{9}{8}%
\left[ \bar{y}^{2}+\frac{\bar{y}}{2\lambda }\right] .
\end{equation}
Its minimum is at $\tilde{\lambda}=\left( \frac{9}{32}\bar{y}\right) ^{\frac{%
1}{3}}$ and
\begin{equation}
\mu \left( \tilde{\lambda}\right) =1.\,\allowbreak 287\,8\bar{y}^{\frac{2}{3}%
}+\frac{9}{8}\bar{y}^{2}-\frac{3}{2}.
\end{equation}
It is evident that there exists a critical radius where $\mu $ turns from
negative to positive. This critical value is located at $\rho
_{c}=1.\,\allowbreak 113\,4$ to be compared with the value $\rho _{c}=1.445$
obtained by B. Allen in \cite{B.Allen}. What is the relation with the large
number of wormholes? As mentioned in I, when the number of wormholes grows,
to keep the coherency assumption valid, the space available for every single
wormhole has to be reduced to avoid overlapping of the wave functions. If we
fix the initial boundary at $R_{\pm }$, then in presence of $N_{w}$
wormholes, it will be reduced to $R_{\pm }/N_{w}$. This means that boundary
conditions are not fixed at infinity, but at a certain finite radius and the
$ADM$ mass term is substituted by the quasilocal energy expression under the
condition of having symmetry with respect to each bifurcation surface. The
effect on the unstable mode is clear: as $N_{w}$ grows, the boundary radius
reduces more and more until it will reach the critical value $\rho _{c}$
below which no negative mode will appear corresponding to a critical
wormholes number $N_{w_{c}}$. To this purpose, suppose to consider $N_{w}$
wormholes and assume that there exists a covering of $\Sigma $ such that $%
\Sigma =\bigcup\limits_{i=1}^{N_{w}}\Sigma _{i}$, with $\Sigma _{i}\cap
\Sigma _{j}=\emptyset $ when $i\neq j$. Each $\Sigma _{i}$ has the topology $%
S^{2}\times R^{1}$ with boundaries $\partial \Sigma _{i}^{\pm }$ with
respect to each bifurcation surface. On each surface $\Sigma _{i}$,
quasilocal energy gives
\begin{equation}
E_{i\text{ }{\rm quasilocal}}=\frac{2}{l_{p}^{2}}\int_{S_{i+}}d^{2}x\sqrt{%
\sigma }\left( k-k^{0}\right) -\frac{2}{l_{p}^{2}}\int_{S_{i-}}d^{2}x\sqrt{%
\sigma }\left( k-k^{0}\right) .
\end{equation}
Thus if we apply the same procedure of the single case on each wormhole, we
obtain
\begin{equation}
E_{i\text{ }{\rm quasilocal}}=l_{p}^{2}\left( E_{i+}-E_{i-}\right)
=l_{p}^{2}\left( r\left[ 1-\left| r,_{y}\right| \right] \right)
_{y=y_{i+}}-l_{p}^{2}\left( r\left[ 1-\left| r,_{y}\right| \right] \right)
_{y=y_{i-}}.
\end{equation}
Note that the total quasilocal energy is zero for boundary conditions
symmetric with respect to {\it each} bifurcation surface $S_{0,i}$. We are
interested in a large number of wormholes, each of them contributing with a
term of the type (\ref{a2}). If the wormholes number is $N_{w}$, we obtain
(semiclassically, i.e., without self-interactions)\footnote{%
Note that at this approximation level, we are in the same situation of a
large collection of N harmonic oscillators whose hamiltonian is
\[
H=\frac{1}{2}\sum_{n\neq 0}^{\infty }\left[ \pi _{n}^{2}+n^{2}\omega
^{2}\phi _{n}^{2}\right] .
\]
}
\begin{equation}
H_{tot}^{N_{w}}=\underbrace{H^{1}+H^{2}+\ldots +H^{N_{w}}}.
\end{equation}
Thus the total energy for the collection is
\[
E_{|2}^{tot}=N_{w}H_{|2}.
\]
The same happens for the trial wave functional which is the product of $%
N_{w} $ t.w.f.. Thus
\[
\Psi _{tot}^{\perp }=\Psi _{1}^{\perp }\otimes \Psi _{2}^{\perp }\otimes
\ldots \ldots \Psi _{N_{w}}^{\perp }={\cal N}\exp N_{w}\left\{ -\frac{1}{%
4l_{p}^{2}}\left[ \left\langle \left( g-\bar{g}\right) K^{-1}\left( g-\bar{g}%
\right) \right\rangle _{x,y}^{\perp }\right] \right\}
\]
\begin{equation}
={\cal N}\exp \left\{ -\frac{1}{4}\left[ \left\langle \left( g-\bar{g}%
\right) K^{-1}\left( g-\bar{g}\right) \right\rangle _{x,y}^{\perp }\right]
\right\} ,
\end{equation}
where we have rescaled the fluctuations $h=g-\bar{g}$ in such a way to
absorb $N_{w}/l_{p}^{2}$. The propagator $K^{\bot }\left( x,x\right) _{iakl}$
is the same one \ for the one wormhole case. Thus, repeating the same steps
of the single wormhole, but in the case of $N_{w}$ wormholes, one gets
\begin{equation}
\Delta E_{N_{w}}\left( x,\Lambda \right) \sim N_{w}^{2}\frac{V}{32\pi ^{2}}%
\Lambda ^{4}x^{2}\ln x,
\end{equation}
where we have defined the usual scale variable $x=3MG/\left(
r_{0}^{3}\Lambda ^{2}\right) $. Then at one loop the cooperative effects of
wormholes behave as one {\it macroscopic single }field multiplied by $%
N_{w}^{2}$, but without the unstable mode. At the minimum, $\bar{x}=e^{-%
\frac{1}{2}}$%
\begin{equation}
\Delta E\left( \bar{x}\right) =-N_{w}^{2}\frac{V}{64\pi ^{2}}\frac{\Lambda
^{4}}{e}.  \label{a5}
\end{equation}

\section{The cosmological constant}

\label{cc}Einstein introduced his cosmological constant $\Lambda _{c}$ in an
attempt to generalize his original field equations. The modified field
equations are
\begin{equation}
R_{\mu \nu }-\frac{1}{2}g_{\mu \nu }R+\Lambda _{c}g_{\mu \nu }=8\pi GT_{\mu
\nu }.
\end{equation}
By redefining
\begin{equation}
T_{tot}^{\mu \nu }\equiv T^{\mu \nu }-\frac{\Lambda _{c}}{8\pi G}g^{\mu \nu
},
\end{equation}
one can regain the original form of the field equations
\begin{equation}
R_{\mu \nu }-\frac{1}{2}g_{\mu \nu }R=8\pi GT_{\mu \nu },
\end{equation}
at the prize of introducing a vacuum energy density and vacuum stress-energy
tensor
\begin{equation}
\rho _{\Lambda }=\frac{\Lambda _{c}}{8\pi G};\qquad T_{\Lambda }^{\mu \nu
}=-\rho _{\Lambda }g^{\mu \nu }.
\end{equation}
If we look at the Hamiltonian in presence of a cosmological term, we have
the expression
\begin{equation}
H=\int_{\Sigma }d^{3}x(N\left( {\cal H}+\rho _{\Lambda }\sqrt{g}\right)
{\cal +}N^{i}{\cal H}_{i})+b.t.,  \label{cc1}
\end{equation}
where ${\cal H}$ is the usual Hamiltonian density defined without a
cosmological term. We know that the effect of vacuum fluctuation is to
inducing a cosmological term. Indeed by looking at Eq. (\ref{a5}), we have
that
\begin{equation}
\frac{\left\langle \Delta H\right\rangle }{V}=-N_{w}^{2}\frac{\Lambda ^{4}}{%
64e\pi ^{2}}.  \label{cc2}
\end{equation}
The WDW equation in presence of a cosmological constant is
\begin{equation}
\left[ G_{ijkl}\pi ^{ij}\pi ^{kl}-\frac{\sqrt{g}}{2\kappa }\left( R-2\Lambda
_{c}\right) \right] \Psi \left[ g_{ij}\right] =0.
\end{equation}
By integrating over the hypersurface $\Sigma $ and looking at the
expectation values, we can write
\begin{equation}
\int_{\Sigma }d^{3}x\left[ \frac{1}{\sqrt{g}}G_{ijkl}\pi ^{ij}\pi ^{kl}-%
\frac{\sqrt{g}}{2\kappa }R\right] =-\frac{\Lambda _{c}}{\kappa }\int_{\Sigma
}d^{3}x\sqrt{g}=-\frac{\Lambda _{c}}{\kappa }V_{c}.  \label{cc3}
\end{equation}
$V_{c}$ is the cosmological volume. The first term of Eq. (\ref{cc3}) is the
same that generates the vacuum fluctuation (\ref{cc2}). Thus, by comparing
the second term of Eq. (\ref{cc3}) with Eq. (\ref{cc2}), we have
\begin{equation}
-\frac{\Lambda _{c}}{\kappa }V_{c}=-N_{w}^{2}\frac{\Lambda ^{4}}{64e\pi ^{2}}%
V_{w}.  \label{cc3a}
\end{equation}
Therefore
\begin{equation}
\Lambda _{c}=N_{w}^{2}\frac{\Lambda ^{4}\kappa }{V_{c}64e\pi ^{2}}V_{w}.
\end{equation}
Since $\kappa =8\pi G=8\pi l_{p}^{2}$ and $\Lambda \rightarrow 1/l_{p}$, we
obtain
\begin{equation}
\Lambda _{c}=N_{w}^{2}\frac{l_{p}^{-2}}{V_{c}8e\pi }V_{w}.
\end{equation}
The cosmological volume has to be rescaled in terms of the wormhole radius,
in such a way to obtain that $V_{c}\rightarrow N_{w}^{3}V_{w}$. This is the
direct consequence of the boundary rescaling, namely $R_{\pm }$ $\rightarrow
$ $R_{\pm }/N_{w}$. Thus
\begin{equation}
\Lambda _{c}=\frac{l_{p}^{-2}}{N_{w}8e\pi }=\tilde{c}\frac{1}{N_{w}l_{p}^{2}}%
.  \label{cc4}
\end{equation}
Due to the uncertainty relation
\begin{equation}
\Delta E\propto \frac{A}{L^{4}}\propto -N_{w}^{2}\frac{V}{64\pi ^{2}}\frac{%
\Lambda ^{4}}{e}\propto A\Lambda ^{4}.
\end{equation}
The fourth power of the cutoff (or the inverse of the fourth power of the
region of dimension L) is a clear signal of a Casimir-like energy generated
by vacuum fluctuations. As a consequence a {\it positive cosmological
constant} is induced by such fluctuations. The probability of this process
in a Euclidean time (not periodically identified) is
\begin{equation}
P\sim \left| e^{-\left( \Delta E\right) \left( \Delta t\right) }\right|
^{2}\sim \left| \exp \left( N_{w}^{2}\frac{\Lambda ^{4}}{e64\pi ^{2}}\right)
\left( V\Delta t\right) \right| ^{2}.
\end{equation}
From Eq. (\ref{cc3a}), we obtain
\begin{equation}
P\sim \left| \exp \left( \frac{\Lambda _{c}}{\kappa }V_{c}\right) \left(
\Delta t\right) \right| ^{2}.
\end{equation}
As an application we consider a periodically identified Euclidean time
\begin{equation}
\Delta t=2\pi \sqrt{\frac{3}{\Lambda }}
\end{equation}
and we admit that a cosmological volume is given by
\begin{equation}
V_{c}=\frac{4\pi }{3}\left( \sqrt{\frac{3}{\Lambda }}\right) ^{3},
\end{equation}
namely
\begin{equation}
\exp \left( 3\pi /l_{p}^{2}\Lambda _{c}\right) .
\end{equation}
Thus we recover the Hawking result about the cosmological constant
approaching zero. Note that the vanishing of $\Lambda _{c}$ is related to
the growing of the holes. Even if this assemblage of coherent wormholes
seems to have the right trend to describe both a spacetime foam and a
quantum gravitational vacuum, we have a lot of problems to solve and other
corrections to include:

\section{Acknowledgments}

I would like to thank Prof. T. Montmerle and Prof. J. Paul who gave me the
opportunity of participating to the Conference. I would like also to thank
Prof. A. Perdichizzi for a partial financial support.

\end{document}